# REVISING THE COSMIC STORY

Gennady Shkliarevsky

**Abstract**:  The essay argues that the standard cosmological model is one-sided and incomplete.  Its recognition of the primacy of radiant energy has no rational justification and empirical verification and, therefore, is arbitrary and subjective.  The subjective and arbitrary choice of this foundational "self-evident truth" renders the entire standard model subjective and arbitrary.  Numerous paradoxes and inconsistencies plague the standard model.  As a result, from the perspective of the standard cosmological model the universe appears, as some cosmologists argue, "absurd" and many of its parts seem inaccessible to our understanding.

The essay attempts to provide a more balanced approach.  It argues that the recognition of equal importance of both radiant and non-radiant forms of energy and matter is essential for a comprehensive and objective understanding of how the universe works.  The essay also tries to outline the new cosmological model that recognizes the equal importance of both types of energy and matter and of their complementary relationship.

**Key words**:  radiant energy, non-radiant energy, black holes, space curvature, the Big Bang, the inflation, conservation, the process of creation.

> There are more things in heaven and earth, Horatio,
> Than are dreamt of in your philosophy.
>
> "Hamlet"
>
> – William Shakespeare

## Introduction

We humans are storytellers by nature.  We need stories to be able to deal with reality.  Stories organize the facts we choose and the way we interpret them.  It does not matter whether we use words, mathematical symbols, physical formulae, sounds, or images, the story remains the main organizing principle.  Our stories are very different but they all have one thing in common:  they all rest on truths that we accept unquestionable as self-evident.  These self-evident truths constitute the foundation for our stories.  It is they that ultimately define the narratives that we construct.

Our universe is one of the subjects about which we write stories.  Decades ago Carl Sagan, perhaps the most celebrated astronomer in history, told us a cosmic story that fascinated millions of readers and viewers.[1]  Sagan's story brimmed with optimism and hope for the future.  It depicted cosmos as a warm and hospitable place full of promise and possibilities—the "harmony of the worlds," as he put it. "Somewhere," Sagan mused, "something incredible is waiting to be known." His story unmistakably conveyed confidence and hope for the future. The enthralling grand cosmic spectacle inspired Sagan. He witnessed for all his readers and viewers a creative process of enormous power and possibilities. This process became implicitly the self-evident truth that organized his entire story.

Much time has passed since then. Today we know a great deal more about our universe than we did back then. Yet despite this knowledge and most likely because of it, our optimism has eroded and our hopes have largely dashed. The more knowledge we have acquired, the more we have come to the realization that there may be much in the universe that we will probably never know; that contrary to Sagan's comforting vision, the universe is a bleak, indifferent, and inhospitable place full of bewildering puzzles and irresolvable paradoxes; and that it offers little promise for the future—in fact, does not promise much of a future at all. The only certainty we now have is that this universe is inevitably doomed to destruction and death.

The dream of "precision cosmology" that many scientists had has given way to new cosmological physics. Its formidable mathematics describes numerous parameters of our universe and their values, but it gives us no sense of the physical meaning of these parameters.[2] The more we know, the less we understand. Michael Turner, a well-known astronomer, has characterized the contemporary state of cosmology as the "absurd universe"; and he is not alone in holding this view.[3]

In their quest for "precision cosmology" modern scientists have lost the inspiring vision of our universe, intuitively grasped by Sagan, as a grand creative enterprise. The process of creation has receded into the background. Somehow, it has evaded the attention of contemporary cosmologists and, as a result, has remained largely unexplored and unexamined. The quest for "precision" has displaced the process of creation the main organizing principle of the story told by the modern cosmology.

The failure to examine and justify its main organizing principles has had a deadening effect on modern cosmological theory. Such failure is not unique to our time and our cosmology. In fact, it has occurred many times in the past. The story with geocentric theory is a good example of the way our science has suffered in the past due to rigidity and the lack of creativity.

New knowledge is a result of new stories we write and new truths that underlie these stories. New truths and stories affect what we observe and how we interpret our observations. However, the old truths often persist and refuse to leave, which stagnates the evolution of our knowledge. The failure to understand how the process of creation works makes people weary of changes. As a result, they resist new ideas and approaches in fear of losing what they have.

This essay represents an attempt to rethink the old truths and revise the old cosmological story. The new story is not to dismiss the old truths. On the contrary, it will re-interpret them and include them into a new and broader frame as its particular cases—i.e. cases that are true under particular circumstances or assumption. In addition to the old truths, the new frame will also include new insights and ideas that have been accumulating in the periphery of mainstream cosmology. In other words, the new story seeks to be more inclusive and comprehensive. It will also make sure that the truth that constitutes the foundation of the new story is objective and can withstand the test of rational justification and empirical verification.

**The "Absurd Universe"**

The standard cosmological model today is, in the opinion of many practitioners, robust.[4] Few would agree today with the sardonic remark attributed to physicist Lev Landau that "cosmologists are often in error though seldom in doubt." Nothing could be further from the truth today. Modern cosmology is a result of many years of meticulous studying of the cosmos. It strives for precision based calculations and observations, not conjecture.

The story that contemporary cosmological theory tells begins about 14 billion years ago in a mysterious event called inflation when space suddenly expanded with incomprehensible speed that exceeded the speed of light many times over. As the modern cosmological thinking goes, at some point our entire universe--all the stars, all the galaxies, everything—was the size of a peach and had a temperature of over quadrillion degrees. A huge thermal explosion of this orange called the Big Bang followed shortly after the inflation. The heat reached unbelievable temperatures that were higher than anything we can imagine. The cooling gave rise to atoms, molecules, stars, galaxies, and much else of what we see today.[5] All the elements that make our universe and us were formed in the span of about a dozen minutes.[6] Not much else is known about the early stage. A dense fog of scattering of energetic particle obscures early developments. All we can see today is the relic of all-pervasive glow produced by the Big Bang that we discovered in the 1960 and that is known as the cosmic microwave background radiation. Whatever forms of matter that existed back then have completely disappeared from view. Only populations of proto-galaxies and primordial starts that emerged several million years after the Big Bang are accessible to our most powerful telescopes.[7]

Amazingly, this fantastical story holds up to all current observations. Astronomers have observed the leftover electromagnetic radiation from the young universe, measured the abundance of the lightest elements and found that they all line up with what the Big Bang predicts and have done much else by way of gathering empirical data. All these date confirm their theory. As far as modern cosmology is concerned, its model is an accurate description of our universe.[8]

The contemporary cosmological story tells much more that what happened in the past and what the current state is. Based on what they know, they predict that the universe that had a beginning will have also an end. Eventually, galaxies will drift apart, stars will explode and die, and planets, including the Earth, will become little more that frozen remnants of what they are now. Following the laws of physics, everything will cool to temperatures as close to absolute zero as quantum theory allows. The universe will reach the state of maximum entropy, or as cosmologists call it the "heat death," when no free energy will be available to do any work.[9] The universe will die. There are, of course different scenarios of this final stage. Some cosmologists see a possibility for a Big Crunch followed by another Big Bang, but whatever the case may be, the universe in which we now live will one day be no more.

The above story of the universe is not the only one available today. There are also some competing narratives, such as the static universe theory or multiverse cosmology that emerged from Everett's "many-worlds interpretation of quantum theory."[10] Then there is also the story of the "ekpyrotic" universe[11] (from the Greek word for "fire") that is the latest iteration of the theory of unending bang-bounce cycles and its extension called cyclic cosmology--both inspired by the string theory.[12] These alternative hypotheses have produced only discord and divisions in the camp of Big Bang cosmologists as they undermine the standard model. While they have their own loyal following, they are hardly in competition with the Big Bang cosmology.[13]

Despite its unquestionable success in dominating the current view of the universe, the standard model of the universe raises numerous questions that it fails to answer. Its explanation of the beginning of the universe does not explain much. If the universe emerged 13.9 billion years ago, where did it come from? The idea that the universe emerged out of nothing is so perilously close to the religious story of creation that some prominent cosmologists, such as Roger Penrose, have attempted to explain the conditions that precede the Big Bang. As Penrose argues: "If we want to know why the universe was initially so very special, in its extraordinary uniformity, we must appeal to completely different arguments from those upon which inflationary cosmology depends."[14] Penrose's argument has been eventually rejected.[15] The "consensus" that has eventually been restored claims that a high entropy (inflation) field came to exist in some way unknown to us, and that the "right values" for the initial conditions and low entropy inflation were due to random fluctuations.[16]

It is worth reminding that inflation is largely an ad hoc idea. It was not a result of some fundamental principles or empirical evidence. Inflation was introduced to account for the extremely uniform distribution of energy/matter and the geometric flatness of space. However, efforts to elaborate this idea have led to theorizing a pre-inflationary phase in which initial quantum fields were highly disordered or entropic. In order to explain the emergence of low entropy state from these chaotic and high entropy conditions, cosmologists have evoked random fluctuations that, in their explanation, led to a "hot" Big Bang; and the subsequent cooling that generated fundamental forces and nucleosynthesis.[17]

As one can see, controversy shrouds the story of the beginning of the universe and creates doubts as to its veracity. If there is a possibility that there may have been no beginning, is it possible that there will be no end? Thus the entire story constructed by Big Bang cosmologists appears to be on a very shaky ground, not the kind of precision theory that it claims to be.

The paradoxes do not end there. Modern cosmology offers no explanations for the most common phenomena we observe in the universe. What is gravitation? Curvature of space is no answer to this question. Theory of relativity merely tells us that the geometry of space depends on the mass of objects, but it says nothing about the physical meaning of such curvature. Black holes, while well documented, still remain a mystery.[18] Modern cosmology can offer little more than new riddles involving dark energy to explain the apparent expansion of the universe. Although cosmologists use the concept of dark matter, they offer little more than promises to explain what this matter actually represents. Finally, modern cosmology has simply lost half of the universe and cannot find it. In accordance with the story we hear from cosmologists, following the Big Bang our universe produced matter and anti-matter in equal amounts. Yet modern cosmology cannot find these potentially vast amounts of anti-matter in our present-day universe.

**Critique of the Current Cosmological Model**

Inclusiveness is the most important criterion for objectivity. An objective description of reality must view an object or a phenomenon from all possible directions and include all possible views and perspectives. Even a cursory glance at contemporary cosmological theory shows that it fails this test of objectivity.

Perhaps the most distinct feature of the standard cosmological model is the primacy of radiant energy. According to the standard model, our universe begins with the huge explosion of radiant energy. It also ends with the disappearance of free radiant energy that can do work. The diffusion of radiant energy, or entropy production, will eventually lead to the so-called thermal death of the universe.

It is common knowledge that radiant energy is not the only form of energy available in our universe. We know that there is also non-radiant energy that we see in gravity, magnetism, dark energy, dark matter, black holes, and other objects and phenomena that are either observed or theorized by cosmologists. Despite the importance of non-radiant energy, in the cosmic story told by the standard model it plays a role that is secondary to that of radiant energy. For example, its part in such important events as the birth and death of our universe is negligible. As Dirac noted, physicists had always arbitrarily ignored the negative energy solutions to his famous equation. Why do contemporary cosmologists consider radiant energy more important than non-radiant? They provide no rational justification or empirical confirmation for the choice they make. One can speculate that the reason for this preference is the fact that radiant energy is more familiar and more intuitively acceptable to researchers, just like the geocentric view of cosmos appealed to ancient cosmologists. The appeal of the familiar and intuitively self-evident cannot be the basis for a theory that claims objectivity. We must consider a possibility that cosmos does not have preference for priority of one type of energy over the other;

both might be equally important. The question then arises, how would the recognition of this possibility affect cosmological theory.

Also, there is another unwarranted preference in the standard cosmological model. It is the preference for the $2^{nd}$ law of thermodynamics. Indeed, this law is one of the fundamental laws of the universe—one of but not the only. There is another law that is at least equally important. It is the law of conservation. The standard cosmological model treats the $2^{nd}$ law of thermodynamics as primary. For example, the Big Bang that plays a central role in the standard model clearly violates the law of conservation. Consequently, contemporary cosmology treats the inflation and the Big Bang as a mystery that may very well be in principle inaccessible to our understanding. Black holes observed and theorized by cosmologists also appear to violate the law of conservation as matter, energy, and information disappear into the black hole without a trace, never to re-emerge again. The dismissive treatment of the law of conservation in the standard model has raised objections from prominent cosmologists. Edward Harrison, for example, emphasized: "The conclusion, whether we like it or not, is obvious: energy in the universe is not conserved."[19] P. J. E. Peebles has reached a similar conclusion. He writes: "The resolution of this apparent paradox [loss of energy inside a commoving ball of the photon gas] is that while energy conservation is a good local concept, . . . there is not a general global energy conservation in general relativity."[20]

The uncritical preoccupation with radiant energy has shaped many aspects of the standard model. It is largely due to this preoccupation that cosmologists favor the theory of the Big Bang as the point of origin for the universe. This theory gained a great deal of popularity when it was first introduced in the 1960s. However, elaborations of this theory have mired it in numerous disputes and controversies.

The main problem is that there is no incontrovertible evidence that the Big Bang actually did take place. In support of the Big Bang the standard model uses the apparent expansion of the universe and the observed cosmic microwave background (CMB) radiation. The observed CMB is, according to the standard model, an "afterglow" from a time about 400,000 years after the Big Bang had supposedly occurred. However, both can hardly be regarded as ironclad proofs. The expansion of the universe is inferred from red-shifted spectra of distant galaxies, not something we observe directly. Also, critics argue that although expanding universe is consistent with the Big Bang, it does not necessarily demand a Big Bang as its cause.[21]

Rather than support the Big Bang, the CMB has generated more controversy and confusion in this theory. For one thing, the argument that the Big Bang with its infinite temperatures is a "proximate cause" of an energy field near absolute zero is really puzzling.[22] In the view of the proponents of the Big Bang, this event was spontaneous, unguided, and undirected. Yet, the CMB shows remarkably uniform distribution. A spontaneous event that could produce such uniformity is extremely unlikely and could be easily suggest fine-tuning and the presence of a Designer. Explanations of this fact have not been successful. Some researchers have tried to argue that although the heat was uneven in the beginning, it radiated and warmed up the rest of the universe. Indeed, radiation spreads in vacuum space very quickly, it still does so at a finite speed—the speed of light. Given the visible size of our universe, some 13.9 billion years since the Big Bang is not enough to equalize temperature in all parts of the cosmos. This is called the horizon problem.[23]

In order to salvage the Big Bang from becoming an argument for creationism, cosmologists came up with the idea of inflation. They offered a speculation that the space remarkably unfolded practically instantaneously at the birth of our universe, which ensured the uniformity of future distribution. However, the solution only complicated the problem even more. For one thing, the only evidence for inflation taking place is the one it is supposed to explain—the CMB. Inflation does not have an independent proof.

But there is more. Inflation was supposed to create an isotropic universe. Indeed it did. The temperature differences in CMB are very small; in fact, they are ten times smaller than the theory predicts, which creates a bit of a problem.[24] Also, the inflation theory predicts similar isotropy on a large scale, not just on a small scale. Yet the evidence from observations of the CMB contradicts this prediction. The high-resolution Planck data show significant variations, including cold spots, on the large scale, which is inconsistent with the inflation theory.[25] These are just a few but by no means all problems with the Big Bang and the inflation. Both theories have become so strange that even secular cosmologists harshly criticize them.[26]

Understanding the physical meaning of cosmological events and facts is another problem are for modern cosmology. Its formidable mathematization tells us how each particle in the universe must behave, but it gives us little sense of the Universe in which we actually live. The "precision cosmology" has realized its dream. It provides the values for the main cosmological parameters, but it does not help us understand the physical sense of the substances to which these parameters refer.[27] The conceptual problems that plague the standard model are a constant source of bitter debates and re-examinations.[28] The new intriguing paradoxes that have been identified make the standard model even more exotic, if not "absurd."[29] Cosmological physics lays out mathematical rules for the universe, but does little to explain the way the universe actually behaves.[30] As Dr. Sean Carroll has explained in an op-ed piece in the New York Times, physicists and cosmologists placed understanding theories in the "too hard" basket."[31] This failure at understanding has prompted Turner's well-known characterization of the standard model as the "absurd universe."[32]

As has already been mentioned, there is much about the universe that the standard model fails to explain. These unexplained phenomena include gravitation and magnetism. In order to explain some gravitational phenomena, the standard model had to postulate the existence of dark matter. They also resorted to dark energy that, in their view, causes the acceleration in the expansion of the universe.[33] According to calculations, dark energy and matter supposedly constitute 95% of the total density of energy and matter. Modern cosmology uses them in its equations, yet they remain nothing more than a hypothesis. The fact that modern physics can say nothing about some of the most important entities in the "consensus universe" increasingly appears scandalous.[34]

Another embarrassment for the standard model is the fact that it lost about half of our universe. According to the standard model, in the wake of the Big Bang the universe had equal amounts of matter and anti-matter. Yet scientists can trace miniscule amounts of anti-matter in our universe today. The problem of the loss of anti-matter remains a persistent mystery; and so do black holes, gravitation, vacuum, and much, much else.

To summarize, the above discussion shows that the basis of the standard cosmological model is subjective, arbitrary, and exclusionary. Its story has roots in the assumption regarding the primacy of one type of energy over the other—an assumption that has no rational justification or empirical verification. No doubt that this subjective foundation affects the entire cosmological theory. Its subjective and exclusionary nature may very well be the source of numerous inconsistencies and paradoxes. They may be the reason why this theory cannot explain many important aspects of our universe. These shortcomings and failures of the dominant cosmological theory clearly show that a more general and comprehensive view is needed to address these concerns.[35] We cannot accept the limits to our knowledge that the standard model as insurmountable.

## The Cosmic Story Revised

### *The Relationship Between Radiant and Non-Radiant Energy*

As has already been pointed out**,** radiant and non-radiant energy are both part of the universe. The universe is the common frame in which these two types of energy co-exist with each other. The standard model has not justified its preference for radiant energy, which suggests that the reason for preference is nothing more than the fact that radiant energy is more familiar and accessible to direct observation. By contrast non-radiant energy can only be observed indirectly, which makes it more difficult to theorize.

Since radiant and non-radiant energy co-exist in one universe, they must be interrelated; and if they are interrelated, there must be some balance between these two types of energy and matter in our universe. Our universe is unique. It is all there is. Nothing can come into it from outside and nothing can disappear from it because there is nowhere to disappear. Consequently, everything must be conserved. The law of conservation must apply to the entire universe and all energy and matter in it—both radiant and non-radiant.

The nature of radiant energy is to radiate. It radiates from regions of high density to regions of low density. When radiant energy radiates away, it does not disappear. It is absorbed elsewhere in the universe. If this is true for radiant energy it must also be true for non-radiant energy, i.e., it must be conserved. Conservation of non-radiant energy, just like conservation of radiant energy, requires movement from areas of high concentration to areas of low concentration.

We know what areas of high concentration of radiant energy look like. We see them in galaxies, stars, and planets. Areas of high concentration of non-radiant energy look very different. Since these areas are dominated by non-radiant energy, they do not emit light. They appear to observers as . . . well, empty space or vacuum. Consequently, when radiant energy radiates away, it leaves behind areas of concentration of non-radiant energy that appear to be empty

Since neither radiant nor non-radiant energy can disappear from the universe, they must coexist; moreover, in accordance with the law of conservation, radiant and non-radiant energy and matter should be in balance. Density must play an important role in regulating their relationship and maintaining them in balance. Areas of high concentration of radiant energy, like active galactic nuclei, eventually give rise to non-radiant energy. What is true for radiant energy must also be true for non-radiant. Therefore, non-radiant energy must also give rise to radiant energy.

How then can non-radiant energy give rise to radiant energy? In order to answer this question, we have to take a close look at phenomena and objects that manifest non-radiant energy. Black holes are perhaps the best-known objects of high concentrations of non-radiant energy.

*Black Holes and Radiant Energy*

As is well known, black holes have no radiant energy. Consequently, they have no charged particles and emit no light. Since black holes have no charged particles, they cannot generate magnetic fields that surround black holes. The source of magnetic fields that surround black holes is moving radiant particles that come from galaxies, and particularly their active galactic nuclei with accretion disks and high-temperature plasma halo that surround black holes.[36]

The magnetic fields and magnetic field lines guide charged particles around the black hole and are believed to be responsible for powerful relativistic jets of plasma, or Gamma-ray bursts, radiated from the coronae.[37] As Darren Dowell, HAWC principal investigator for NASA's Jet Propulsion Laboratory in Pasadena, California, explains: "The spiral shape of the magnetic field channels the gas into an orbit around the black hole."[38] Yigit Dallilar, a scientist from the University of Florida, confirms: "The required magnetic field is either accumulated from surroundings via accreting material, or simply generated within the accretion flow." In other words, the environment around the black hole, not the black hole itself, governs magnetic fields that surround the black hole.[39]

Thus, high density of radiant energy and matter in the AGN create high temperature and pressure that energize particles. This radiant energy is incompatible with black holes that are objects of non-radiant energy and matter. For this reason, the radiant energy generated in the area of high density around the black hole is not absorbed into the black hole. It is radiated away into areas of low density of radiant energy and matter.[40]

Therefore, the removal of radiant energy does not leave the space it occupied empty. When radiant energy is gone, the space it has occupied becomes dominated by non-radiant energy. Black holes are domains of non-radiant energy.

*Quantum Vacuum and Non-Radiant Particle Pairs*

But what exactly occupies this space? What does it consist of? Contemporary cosmology does not provide a clear answer. It recognizes the fact that vacuum is not empty. However, it does not really tell us what fills it. Cosmologists contend that vacuum is filled by something they call "virtual particles" and leave it at that. The term "virtual particle" is an ambiguous one. It suggests that these particles are not like real particles. However, mainstream scientists do not tell us what exactly these "virtual particles" are. They tell us that "virtual particles" have energy—real energy. If these particles have real energy, they must be real, not some "virtual" variety, which is neither here nor there. This mental gymnastics merely tries to explain things away, not provide a clear answer.

When scientists bombard vacuum with high energy of no less than 1.022 MeV, they obtain two real particles—an electron and a positron, or matter and anti-matter. The process is well known as "pair production." When these particles appear, they do not hang around for too long. As matter and anti-matter, they clash an disappear in the process that is misleadingly called "annihilation." The "annihilation" releases the amount of energy equal to what was required for their production. The release of energy produces another pair that in turn also "annihilate" each other and so on.

The law of conservation tells us that nothing can appear from nothing and nothing can disappear without a trace from our universe, we have to conclude that the particle pair must have real existence; and the 1.022 or more MeV merely make the pair appear as radiant particles in our three-dimensional space. If these particles have existed in some undetectable form prior to their appearance, they must have some real characteristics that we simply do not have the capacity to observe directly. However, the appearance of the two particles as a pair allows one to draw some conclusions about how they were prior to their appearance:

1. Since the particles appear as a pair, one can logically conclude that in whatever form they exist prior to their production, they must exist as a pair, not as two individual particles. William Straub theorizes them as "four-component object consisting of two stacked spinors," or what Dirac called "bispinor."[41] In their generalization of Dirac equation, Hossain Javadi, of the Islamic Azad University, and his co-authors, also theorize the existence of a non-radiant pair that is converted into two fermions by 1.022 MeV of energy.[42] Don Hotson devoted his two-part article "Dirac's Equation and the Sea of Negative Energy" to the discussion of particle pairs that fill vacuum space.[43]

2. Non-radiant particles do not have a charge, but they do have a spin.

3. Since they are particles, they also have energy. This energy can only be the energy of their spin.

4. Since a non-radiant pair does not have a charge, it cannot have more than two dimensions. The size of the pair must also be well within the Planck range.[44] These two characteristics make direct observation of the pair in non-radiant mode impossible.

5. Since electron and positron have opposite spins, each particle in a non-radiant pair also must have opposite spins, i.e., their spin vectors should point in different directions. However, the spins must be conjugate; otherwise, they would interfere with each other. Since the two particles have different spin directions, they also have different space orientation vis-à-vis each other.

Space transmits energy in all frequencies. Therefore, particle pairs should be capable of oscillating at different allowable frequencies. No particle pair can oscillate at all frequencies because that would contradict the law of conservation and the Pauli exclusion principle. Therefore space must be made up of packets that include many non-radiant particle pairs that fill all available energy levels. Simple calculations that use the periodic table show that radiant elementary particles can vibrate in 136 different ways. If this is true for radiant particles, the same must be true for non-radiant ones.[45] Therefore, packets should consist of 136 non-radiant particles pairs packed matryoshka style inside each other.

All pairs in a packet are oscillating harmonically in accordance with the law of conservation.[46] They all spin at the speed of light; their frequencies vary according to their size. Particle pairs are limited in size; they cannot be smaller than allowable limit. Again, the law of conservation and the Pauli exclusion principle require this limit. All particle pairs packed inside each other move at the speed of light. No two pairs can occupy the same energy level due to the exclusion principle. For this reason, as particle pairs are packed, each pair should occupy a different energy level. Their frequency (energy level) depends on their size: the smaller the size, the higher the frequency, or energy level. But there is one limitation. The reduction in size is not infinite. When the limit is reached and the size can not be further reduced, the only way to increase the energy level is to increase the speed of rotation, that is, to go faster than the speed of light, which is not allowed. When this limit is reached, a packet is full and can take no more energy increases.

These packets of non-radiant particle pairs are capable of transmitting radiant energy at all available levels. Such packets constitute vacuum; they make up the space.

*Interactions Between Radiant Particles and Packets of Non-Radiant Particle Pairs*

As has been stressed earlier, the nature of radiant energy is to radiate. By radiating, radiant energy conserves itself. It radiates from areas of high density to areas of low density. By transmitting radiant energy from one region to another, or from the point of emission to the point of absorption, packets of non-radiant particle pairs make conservation of radiant energy and matter possible. That is why packets of non-radiant particle pairs and radiant particles must interact.

One form of their interaction is the transmission of radiant energy from the point of emission to the point of absorption. A bare radiant charge polarizes vacuum. When a source emits a radiant particle, it is immediately surrounded by strings of polarized non-radiant particle pairs. Spin plays a critical role in connecting a radiant and non-radiant particle. In order for the radiant particle and non-radiant particle to interact, their spin vectors, just like spin vectors in a particle pair, should point in different directions. Similarly, they should have different orientation in space vis-à-vis each other. If their spins/vectors point in the same direction, they will interfere with each other and will not be able to establish a connection that can transmit energy.

When a source emits a radiant particle, this particle perturbs the vacuum that tries to maintain its integrity and its temperature near absolute zero. Strings of polarized packets of non-radiant particle pairs surround the perturbing radiant particle. One end of these strings surround the energy particle at the point of emission while another end is connected via spin to the point that has a potential for absorbing this radiant energy. When the string of polarized packets of non-radiant particle pairs connects the point of emission with the point of absorption, the string transmits energy "bucket-brigade" style at velocity c, thus dumping the charge and conserving the vacuum space.

Since packets of non-radiant particle pairs make up the vacuum space, they are the only candidates to account for other manifestations of non-radiant energy in vacuum space, such as gravitation and magnetism. Polarized strings of packets that converge on radiant objects in space make up gravitational lines that constitute gravitational fields. Convergence of gravitational lines on radiant objects creates a gradient of transverse space compression. Such compression has the effect of space curvature theorized by Einstein. The capacity to compress space is the source of gravity. This transverse compression results in increased density of gravitational lines and therefore obeys the inverse square law. Under transverse compression, cross sections of the field closer to the object will have a greater density of gravitational lines in the inverse square proportion. Finally, I would add that polarized packets of non-radiant particle pairs also constitute magnetic lines that form magnetic fields. All the above examples manifest different ways in which packets of non-radiant particle pairs interact with radiant particles and masses.

*Black Holes, Magnetic Fields, and Space Curvature*

Just like any other object, black holes compress space. The only difference is that black holes compress space to the extreme.

In contrast to radiant objects where gravitational lines of polarized packets of non-radiant particle pairs attach themselves to radiant particles, in the case of a black hole they attach via spin to packets of non-radiant particle pairs that are already packed into the black hole. The compression in this case is extreme since there are no radiant particles that spread out gravitational lines and non-radiant particle pairs use spin energy to attach directly to each other.

The gravitating non-radiant vacuum space comprised of incoming gravitational lines of non-radiant packets is wraps around the black hole following its geometry. The packing is extremely tight. A popular description is that black holes "curve" or "fold" the space "onto itself." The description is metaphorical. The physical meaning of this metaphor is one of tightly compressed packets of non-radiant particle pairs. A reflection by Vasily Komarov helps understand the connection between non-radiant particles and black holes:

> . . . [The] quantum spin of fundamental particles is closer related to the mechanical angular momentum than you think and therefore it is closer related to the speed of light. This relationship is a bit more complicated than it sounds as well as the same surprisingly accurate relationship between rotation speed of black holes and speed of light. Roughly speaking, for fundamental particles probably it is the result of vibrations and beating of several frequencies.[47]

As has already been mentioned, black holes are surrounded by magnetic fields that suspend them in space. Magnetic fields also constrain black holes to their size.[48] Since a black hole contains no particles with radiant energy, it cannot generate the magnetic fields that surround it. The source of these fields is moving radiant particles that come from the galaxy and reach extreme density in the active galactic nucleus (AGN), the accretion disk, and the high-energy plasma halo. The lines of the magnetic fields guide charged particles around the black hole. They do not and cannot guide them across the event horizon and into the inner regions of the black hole as there are no and cannot be any magnetic fields inside black holes.

The high density of radiant energy and matter in the AGN create high temperature and pressure that energize radiant particles. These particles generate radiant energy that is radiated away into space and regions of low density of radiant energy. The black hole cannot absorb these highly energized particles since such particles will destabilize the interior of the black hole and violate its integrity. When all radiant energy leaves the space around the black hole, the remaining vacuum space now has exclusively non-radiant forms of energy and matter. The space consists of tightly compressed packets of non-radiant energy and matter. This non-radiant energy and matter cannot destabilize the black hole and violate its integrity.

Black holes have a bad reputation among modern cosmologists. They are bad guys of our world that supposedly devour anything that crosses their event horizon; everything is subject to destruction and most importantly all the information that is encoded in radiant objects and energy. Many contemporary cosmologists believe that the information about matter that disappears into black holes is lost forever.[49] Nothing can be further from the truth. Black holes do not devour everything that comes close to them. Ethan Siegel compares them to Cookie Monster, scattering matter around rather than destroying it. He writes:

> It's estimated that, contrary to the popular picture, upwards of 90% of in-falling matter will never make it inside a black hole at all. Instead, it's spewed back out into the outer regions of the galaxy, where it can fuel the formation of new stars and return to the interstellar medium once again.[50]

Shep Doeleman, director of the Event Horizon Telescope (EHT) also comments:

> Black holes have this intense gravity, but they're trying to squeeze everything into this small volume . . . There are not very many ways to do that. You compress gas and it heats up. Take all this gas and it's streaming toward the event horizon [the boundary of a black hole] and getting closer and closer together, it gets hotter and hotter and wants to fly away. Convincing all that gas to go through the event horizon is not a straightforward thing.[51]

Many researchers find the loss of information paradoxical since entropy cannot destroy information. Such destruction would contradict the $2^{nd}$ law of thermodynamics.[52] Stephen Hawking envisioned a possibility that some information may actually escape destruction. In his view, quantum vacuum fluctuations convert into pairs of particles. One of these particles crosses the event horizon and disappears forever, while the other particle escapes into space in the form of radiation called "Hawking radiation." As particles escape, they carry with them some information that, consequently, is conserved.[53] Physicist Lee Smolin, of the University of Waterloo's Perimeter Institute for Theoretical Physics (PITP), thinks that black holes contain the seeds of new universes.[54]

There are quite a few facts that contradict the view of black holes as all consuming.[55] This work suggests that non-radiant particle pairs that are packed into black holes carry with them very important information on the speed of light, spin, particles' orientation in space vis-à-vis each other, and levels of energy that particles can occupy. The law of conservation and the law of entropy are valid for the entire universe—that is, both its non-radiant and radiant sphere.

Magnetic fields play a critical role in sustaining black holes.[56] They separate the black hole from radiant matter and energy, thus maintaining the integrity of black hole structure. As have already been mentioned, they also constrain the size of the black holes. The result is the tight compression and packing of space inside black holes.[57]

When all radiant energy from the space around the black hole disappears, so do magnetic fields. Without the constraint that magnetic fields provide, the black hole unravels. It can no longer compress the space and the space that was packed inside the black hole begins to unfold. This unfolding is the closest physical equivalent to what modern cosmologists call metaphorically "inflation of space."[58]

*The Rise of Radiant Matter in Non-Radiant Vacuum Space*

Black holes conserve packets of non-radiant particle pairs. When a black hole disappears and can no longer conserve particle pairs. Conservation requires resources. Therefore, packets of non-radiant particle pairs must get access to new resources that would ensure their conservation.

There is only one resource available to particle pairs in the vacuum space. They have only each other as a resource that can be used for conservation. In order to gain access to this source, packets have to organize into stable units. Nature has many examples when entities conserve themselves by aggregating into larger units. Particles with spins are no exception.[59] Natural selection favors such form of conservation. Aggregation and self-organization makes possible for packets of non-radiant particle pairs to combine their energies, thus acquiring additional resources that can be used for conservation.

Since spin is the only source of energy available to packets of non-radiant particle pairs, that is the resource they use in creating their combinations. In order to create combinations, spins should be conjugating, that is, their spin vectors should point in different directions. Such conjugal spins will not interfere with and will actually complement each other.

The packets of non-radiant particle pairs are identical. But in order to be conjugal, they must have different orientation is space vis-à-vis each other. In order to combine two packets, they should have different orientation vis-à-vis each other: the down spin of one packet correlates with the up spin of another packet. Just like quarks, they know their left from right.[60]

A combination of two packets is relatively unstable since it can take another packet into its combination. The third component can be either spin down spin up. If the third packet is spin up, we get a combination of two spins up and one spin down. This combination is very familiar to physicists as it represents the organization of a proton. This combination is relatively stable.

If the third packet is spin down, then we have the combination of two down spin components and one up spin component. This arrangement is also familiar to physicists. It is the structure of a neutron. As is well known, neutron is relatively unstable particle; it decays and this decay takes 14.5 minutes—a long time for a sub-atomic particle. Physicists know this process as beta decay. The cause of beta-decay is still a mystery.[61] One can, however, venture a conjecture. If the reason for combining is to gain access to additional resources, then the combination of three packets should provide extra energy. However, since all non-radiant energy levels are already occupied, this extra energy cannot be accommodated in the non-radiant mode. The only way to accommodate the increase in energy is to convert this energy into the radiant mode. As a result, one particle pair acquires a charge.

The particle in the pair with the outflow spin is expelled from the combination and forms a three-dimensional shell around it. By convention we designate its radiant charge as -1 and we call it electron. The other particle of the pair with the inflow spin remains in the combination and gives it a total charge of +1. Thus, as a result of decay we get a nucleus with a charge of +1 and the surrounding shell formed by a particle that has a charge of -1.

One can easily recognize in this new arrangement the structure of the simplest atom of hydrogen with positively charged nucleus (proton) and a negatively charged shell around it formed by an electron that carries a charge of -1. This atom is the most common element in our universe. Hydrogen is estimated to account for 90% of all atoms in the universe; and that includes ourselves since close to two-thirds of the atoms in our bodies are hydrogen.[62]

The emergence of radiant energy transforms the three components in the threesome combination into what physicists call quarks. Each quark has a spin of ½ that is either up or down. Unlike spins neutralize each other, so the total spin of the entire combination is always ½. The down quark, takes the charge of – 1/3 and each up quark) take the charge of +2/3. Thus the total charge of neutron equals to  -1/3 plus another -1/3 plus +2/3 charge for a total of 0, which is the empirically verified charge of a neutron. The proton has two up and one down quark and its total charge comes to:  +2/3 plus another + 2/3 plus -1/3 for a total of +1 charge.

Several considerations support this version of the emergence of radiant energy. First of all, it is the observed abundance of nucleons in our universe.[63] Nucleons do not appear to be fundamental. They have a relatively complex structure that includes parts. Obviously, they are not as fundamental as electron that has a much simpler structure. Also, they are much heavier than electron. If nucleons were made in a way that compounded smaller units into larger ones, we should expect electrons to be far more abundant in the universe than nucleons. Simple hydrogen is about 1012 times more abundant than compound uranium.[64]

Yet in contradiction to this expectation compound nucleons appear to be twice as abundant as the simple electron. While the most common element, hydrogen, contains no neutrons, there is an excess of neutrons in the more complex elements. If you count the unknown but probably large number of neutron stars, there seem to be nearly as many neutrons as protons. Thus there appear to be nearly twice as many nucleons as electrons.[65] The abundance of nucleons is more consistent with the above description of the rise of radiant energy, nucleons, and atoms, rather than the cooling model of nucleosynthesis under the Big Bang theory. Also, the current vacuum temperature of 2.75 K is much closer to the non-radiant spectrum with its temperature close to absolute zero than to infinite temperatures of the hot Big Bang.[66]

The two radiant particles that emerge as a result of beta decay are distanced from each other but they are in contact. The space between the negatively charged electron shell and the nucleus with the positively charged particle (positron) is not empty. It is filled with polarized strings of non-radiant packets that connect the positively charged nucleus with the electron shell and that transmit radiant energy between the nucleus and the shell.

On the outside of the atom, polarized packets of particle pairs surround the shell created by the electron, thus isolating the atom, ensuring its stability, and preventing the charge to disturb the integrity of vacuum space. The strings of polarized packets of non-radiant particle pairs also transmit radiant energy from the point of emission to the point of absorption. Finally, as has been explained earlier, gravity, mass, and magnetism—all owe their existence to the strings of polarized packets of non-radiant particle pairs.

Once simple hydrogen atoms emerge, gravity and electro-magnetic forces bring atoms together. They create huge clouds of aggregated atoms that are further compressed by forces of gravity and electro-magnetism. The effect is the increase in density of radiant energy and matter. This process takes us to the emergence of stars, galaxies, planets and all other "miracles" of the universe.

Galaxies that eventually emerge through the process of self-organization that involves gravity eventually form a tight structure with high density of radiant energy and matter in its center (AGN). As a result of high density, huge amounts of radiant energy are radiated into space, freeing space for non-radiant energy—the process that eventually leads to the formation of black holes. Thus, the entire process that this article has described enters a new cycle.

**Conclusion: A New Cosmology**

There are many stories about our universe. They are all very different. Yet despite these differences, they are all interrelated. These stories are all part of a larger story that is still being written. They are part of our human world that, like the physical universe, is unique. It is all we humans have. Nothing can come to it from outside and nothing can disappear from it. Everything must be conserved.

The story told in these pages does not destroy what has come before it. It merely provides a broader frame that includes everything that has preceded it. It tells a story of the cosmos and our civilization, that is the human universe. It is a story about the eternal process of creation that sustains cosmos and our human world.

The process of creation is central to this story. It provides the basis for the new cosmological model. In contrast to the standard cosmological model that offers no rational justification or empirical verification for its "self-evident" truths, the foundation of the new model can stand the test of rational justification and empirical verification. Without the process that creates new constructs in our mind, we would not be able to perceive and interpret reality. Indeed, to paraphrase the famous adage, there would be nothing rather than something. We can observe the results of the process of creation all around us in the universe: in galaxies, stars, and planets, in the emergence of life, in the rise of the human race and its civilization. We can observe creation in works of art, literature, music, science, and much else. The process of creation makes all our creations possible.

Conservation is essential for the process of creation. Creation is not a whim, not a caprice, and not an accident. It is a result of the meticulous process that brings different entities into a common frame in which they sustain and thus conserve each other.

Conservation requires new resources. Resources are always scarce. Only new and increasingly more powerful levels of organization can provide access to new resources. The $2^{nd}$ law of thermodynamics plays a critical role in this process. It allows various forms of matter and energy to reach equilibrium and thus attain a new and more powerful level of organization that regulates their relationship. Thus our universe is constantly in the process of creation and this process sustains it. The universe has, no beginning and, consequently, no end. It is constantly emerging and thus conserving itself.

The vision outlined in this essay is broader and more comprehensive than the current cosmological model. It is not based on limited truth that arbitrarily assigns primacy to one type of energy. It argues that radiant and non-radiant energy are equal partners and that their partnership is essential for sustaining our universe. These two types of energy are very different from each other and may even seem incompatible. Yet the universe provides a common frame that includes both as particular cases.

As this article has shown, the equilibration of radiant energy and matter gives rise to non-radiant forms of energy and matter; just as the equilibration of non-radiant energy and matter gives rise to radiant forms. Both types of energy transform into each other and this reciprocal transformation ensures the sustainability of the universe.

The cosmological story does not end here. The equilibration of the radiant sphere produces more than galaxies, stars, and planets. It also gives rise to life and consciousness. The emergence of life and consciousness opens the universe up to the creation of infinite number of new and increasingly more powerful levels of organization. It makes our universe truly infinite and ever changing.

In order to include different entities into a common frame, the process of creation must retain and conserve their autonomy. The fact that the process of creation combines differences, retains their autonomy, and thus conserves them has profound implications.

Humans are a product of the process of creation. They have inherited this process and its properties in the course of the evolution. These properties become part and parcel of our world. Our mind can create an infinite number of new and increasingly more powerful levels of organization that open for us access to new resources and most importantly to new perceptions and interpretations of reality. This capacity to acquire new ways of perceiving and interpreting reality is the source of our knowledge.

In order to create new knowledge, we must combine differences, which means that we have to recognize the autonomy of these differences. The recognition of autonomy—both one's own and that of others—is the basis of moral sentiment and morality. Combining different entities and their functions without violating their autonomy conserves them. In order to conserve a function, this function must be enacted. Action is a way of conserving a function. In a way, it is a form of gratifying this function. By acting out who they are, humans gratify their human functions and such gratification is the source of pleasure and aesthetic sentiment. We can agree with Carl Sagan who saw our universe as a warm and hospitable place, as the source of our knowledge and our moral and aesthetic values. In this sense, we can say that the universal process of creation that sustains our universe is the source of what we value: knowledge, justice, and beauty

The new cosmological model outlined in this article in a way revives the vision projected by Carl Sagan. It translates into rational arguments what Sagan grasped intuitively. This rationalization does not deny Sagan's intuitions. On the contrary, it makes them more tangible and etches them into our consciousness better. It, in fact, conserves his intuition. Our consciousness represents the most powerful form of organization of matter. It can create an infinite number of new and increasingly more powerful levels of organization of reality. Through our rational understanding of the process of creation we gain control over this process, thus making the process of creation more efficient. By gaining control over the process of creation, we fulfill this process and open our universe to infinite and eternal creation.

**Notes**

[1] In 1980 Sagan presented to the public his highly acclaimed series "Cosmos: A Personal Voyage." He also subsequently published a companion to his series (Carl Sagan, *Cosmos* [New York: Ballantine Books, 1980]).

[2] P.J.E. Peebles, "From precision cosmology to accurate cosmology," astro-ph/0208037, 2002, https://arxiv.org/abs/astro-ph/0208037; Michael Turner, "Making sense of the new cosmology," astro-ph/0202008, 2002, https://arxiv.org/abs/astro-ph/0202008.

[3] Michael Turner, "Absurd universe," *Astronomy*, v.31, no.11 (2003), p. 44.

[4] Roger Blandford, Jo Dunkley, Carlos Frenk, Ofer Lahav, and Alice Shapley, "Coming of Age of the Standard Model," February 28, 2020. https://doi.org/10.1038/s41550-020-1012-8.

[5] Paul Sutter, "What Happened before the Big Bang?" *Livescience.Com*, April 21, 2020. https://www.livescience.com/what-came-before-big-bang.html.

[6] Paul Sutter, "What Happened before the Big Bang?"